\documentclass[a4paper,11pt]{article}
\usepackage{pos}

\newcommand{\mrm}[1]{_{\rm #1}}
\renewcommand{\d}{{\rm d}}

\title{Dark radiation constraints on light primordial black holes}
\author*[a]{J\'er\'emy Auffinger}
\author[a,b,c]{Alexandre Arbey}
\author[d]{Pearl Sandick}
\author[d]{Barmak Shams Es Haghi}
\author[e]{Kuver Sinha}

\affiliation[a]{Univ. Lyon, Univ. Claude Bernard Lyon 1, CNRS/IN2P3, IP2I Lyon, UMR 5822, F-69622, Villeurbanne, France}

\affiliation[b]{Theoretical Physics Department, CERN, CH-1211 Geneva 23, Switzerland}

\affiliation[c]{Institut Universitaire de France (IUF), 103 boulevard Saint-Michel, 75005 Paris, France}

\affiliation[d]{Department of Physics and Astronomy, University of Utah, Salt Lake City, UT 84112, USA}

\affiliation[e]{Department of Physics and Astronomy, University of Oklahoma, Norman, OK 73019, USA}

\emailAdd{j.auffinger@ipnl.in2p3.fr}
\emailAdd{alexandre.arbey@ens-lyon.fr}
\emailAdd{sandick@physics.utah.edu}
\emailAdd{shams@physics.utah.edu}
\emailAdd{kuver.sinha@gmail.com}

\abstract{Light black holes could have formed in the very early universe through the collapse of large primordial density fluctuations. These primordial black holes (PBHs), if light enough, would have evaporated by now because of the emission of Hawking radiation; thus they could not represent a sizable fraction of dark matter today. However, they could have left imprints in the early cosmological epochs. We will discuss the impact of massless graviton emission by (rotating) PBHs before the onset of big bang nucleosynthesis (BBN) and conclude that this contribution to dark radiation is constrained by the cosmic microwave background (CMB) (with the future CMB Stage 4) and BBN in the lighter portion of the PBH mass range, under the hypothesis that they dominated the energy density of the universe.}

\FullConference{%
  *** The European Physical Society Conference on High Energy Physics (EPS-HEP2021), ***\\
  *** 26-30 July 2021 ***\\
  *** Online conference, jointly organized by Universität Hamburg and the research center DESY ***
}

\begin{document}

{CERN-TH-2021-169}

\maketitle

\section{Introduction}

Primordial black holes (PBHs) are a compelling candidate for the dark matter (DM) --- for recent reviews, see~\cite{Carr:2020xqk,Green:2020jor}. A variety of cosmological models predict the formation of PBHs in the early universe, when primordial density fluctuations re-enter the Hubble horizon and subsequently collapse (see~\cite{Carr:2020gox} for a review); thus the mass of PBHs is linked to the size of the horizon at re-entry with\footnote{Throughout the paper, we use the natural system of units with $c = \hbar = G = k\mrm{B} = 1$.} $M \approx 10^{15}\,{\rm g}\times (\dfrac{t}{10^{-23}\,s})$. The PBH initial mass can in principle go down to the Planck scale, at which we do not know the behavior of general relativity. However, observation of the CMB by Planck has put constraints on the Hubble scale during inflation, which in turn translates as a lower bound on the possible PBH mass $M \gtrsim 0.07\,$g~\cite{Auffinger:2020afu}, but this is inflation model-dependent.

In this study~\cite{Arbey:2021ysg}, we aim at constraining the abundance of very light PBHs of mass $M = 10^{-5}-10^9\,$g in a scenario where they come to dominate the energy density of the universe before evaporation and emit massless gravitons as part of the Hawking radiation. For this, we compute the total graviton emission of PBHs in various scenarii corresponding to different PBH angular momentum distributions: PBHs that acquire spin through hierarchical mergers~\cite{Fishbach:2017dwv} or PBHs formed initially with very high spin during an early matter dominated era (EMDE)~\cite{Harada:2017fjm}, which is the main focus of this study. The evaporating PBH density is mostly converted into Standard Model (SM) radiation, thus reheating the universe to a temperature $\rho\mrm{PBH}(t\mrm{RH}) \approx \rho\mrm{SM}(t\mrm{RH}) \propto T\mrm{RH}^4$. Gravitons instead behave as dark radiation (DR), which is constrained by current CMB observations and deductions from big bang nucleosynthesis (BBN) rates. We use these, as well as the promising future CMB stage 4 (CMB-S4) prospects~\cite{CMB-S4:2016ple}, to derive constraints on the spin distribution scenarii.

This range of PBH masses had until recently evaded constraints because there were no measurable imprint of their evanescent presence at such an early epoch. Indeed, PBHs with mass $M \lesssim 10^9\,$g evaporate just before the onset of BBN. PBHs abundance above this mass is tightly constrained\cite{Carr:2020gox}. Refs.~\cite{Papanikolaou:2020qtd,Domenech:2020ssp} nevertheless proposed to use present or future interferometers to measure the scalar-induced gravitational waves generated by the density fluctuations accompanying PBHs. Other constraints may come from the emission of warm DM as part of the Hawking spectrum (see~\cite{Masina:2020xhk,Auffinger:2020afu,Masina:2021zpu} and references therein).

\section{Hawking radiation of rotating black holes}

Hawking has shown that black holes (BHs) are not eternal~\cite{Hawking:1974rv,Hawking:1975iha}. They emit continuous radiation of all (beyond the) Standard Model (SM) particles $i$, with a rate per unit time and energy that depends on their mass and angular momentum
\begin{equation}
\dfrac{\d^2 N_i}{\d t\d E} = \dfrac{g_i}{2\pi} \sum_{l,m} \dfrac{\Gamma_{l,m}^{s_i}(E,M,a^*)}{e^{E^\prime/T} - (-1)^{2s_i}}\,,\qquad \text{with} \qquad T(M,a^*) \equiv \dfrac{1}{2\pi}\left( \dfrac{r_+ - M}{r_+^2 + a^{*2}M^2} \right),
\end{equation}
where $s_i$ is the particle spin, $l = s_i, s_i+1,...$ is the angular momentum number with projection $m = -l,...,l$ and $g_i$ is the multiplicity of particles/antiparticles, as well as the helicity and color. The energy $E$ is corrected for horizon rotation into $E^\prime$ and has a cutoff $E > \mu_i$ at the particle rest mass. $T$ is the Hawking temperature of a rotating PBH. The last factor is $\Gamma_{l,m}^{s_i}$ which is called the ``greybody'' factor and encodes the probability that a particle created at the PBH horizon escapes to spatial infinity (correction to the pure blackbody radiation). Once the rates of emission of all the particles are known, the PBH mass and spin evolutions are deduced through integration over the emitted energy and angular momentum, with definition of the Page factors $f,g(M,a^*)$~\cite{Page:1976df,Dong:2015yjs}
\begin{equation}
	\dfrac{\d M}{\d t} = -\dfrac{f(M,a^*)}{M^2}\,, \qquad \dfrac{\d a^*}{\d t} = a^*\dfrac{2f(M,a^*) - g(M,a^*)}{M^3}\,.
\end{equation}
The effect of rotation on PBHs is to enhance the emissivity of the higher spin particles (such as the graviton), through a complex coupling between the particle's and the PBH's spins (\textit{e.g.}~\cite{Dong:2015yjs}). This causes the PBH to rapidly loose its angular momentum and to evaporate faster than a non-rotating PBH of the same mass. The difference in lifetime is less than an order of magnitude even for near-extremal values of the spin~\cite{Arbey:2019jmj}. Hereafter, we consider that PBHs emit only the SM particles, and the massless graviton of spin 2 with 2 polarization degrees of freedom. The Hawking radiation emission rates are computed with the public code \texttt{BlackHawk}~\cite{Arbey:2019mbc,Arbey:2021mbl}. Substantial details about the numerical methods of computation are described in the \texttt{BlackHawk} manual.

The density of gravitons (hereafter indexed by DR for ``dark radiation'') and the total density of SM particles (sum over all the SM emissivities) at evaporation are
\begin{equation}
	\rho\mrm{DR/SM}(t\mrm{RH}) = \int_0^1 \d a^*\,\dfrac{\d n}{\d a^*} \int_0^{t\mrm{RH}} \d t \int_0^{+\infty}\d E\,E\dfrac{\d^2 N\mrm{DR/SM}}{\d t\d E}\,, \quad \text{with} \quad \int_0^1 \d a^*\,\dfrac{\d n}{\d a^*} = 1\,.
\end{equation}
We have considered a monochromatic mass distribution for simplicity, corresponding to an instantaneous formation of PBHs. The extended spin distribution $\d n/\d a^*$ results in a spread of evaporation times $\tau(a^*)$. Hence, we must be careful when defining the reheating time $t\mrm{RH}$. Either we consider that it is the longest lifetime of a PBH of the distribution (lowest initial spin) $t\mrm{RH} = \max(\tau(a^*))$, or we compute the weighted average of lifetimes
\begin{equation}
	t\mrm{RH} \equiv \int_0^1 \tau(a^*)\dfrac{\d n}{\d a^*}\,\d a^*\,,
\end{equation}
which we believe is more realistic. PBHs effectively loose most of their mass at the very end of their lifetime $\tau$~\cite{Arbey:2019jmj}, thus we can consider that this reheating is instantaneous as a first approximation.

\section{Dark radiation constraints}

The graviton density at reheating is translated into a DR constraint as an extra contribution to the effective number of relativistic neutrino degrees of freedom $N\mrm{eff} = N_\nu + \Delta N\mrm{eff}$ where the fidutial value is $N_\nu = 3.046$~\cite{Planck:2018vyg}. $\Delta N\mrm{eff}$ is usually expressed at matter-radiation equality through
\begin{equation}
	\Delta N\mrm{eff} = \dfrac{\rho\mrm{DR}(t\mrm{RH})}{\rho\mrm{SM}(t\mrm{RH})}\dfrac{g_*(T\mrm{RH})}{g_*(T\mrm{EQ})}\left( \dfrac{g_{*,S}(T\mrm{EQ})}{g_{*,S}(T\mrm{RH})} \right)^{4/3}\left( N_\nu + \dfrac{8}{7}\left( \dfrac{11}{4} \right)^{4/3} \right),
\end{equation}
where the ratio of densities at equality is obtained from the ratio at reheating from scaling relations due to the dilution during expansion and $g_{*}$ ($g_{*,S}$) counts the number of relativistic (entropy) degrees of freedom. These precisely computed functions are available in the public code \texttt{SuperIso Relic}~\cite{Arbey:2009gu,Arbey:2018msw}. The quantity $\Delta N\mrm{eff}$ is constrained by CMB observations with different strictness depending on the electromagnetic modes examined (TT,TE,EE+low E). It will be probed more deeply by the future CMB-S4 experiment~\cite{CMB-S4:2016ple} up to a sensitivity of $\Delta N\mrm{eff} \lesssim 0.023$. on the other hand, a non-0 $\Delta N\mrm{eff}$ contribution modifies the thermal history of the universe and thus the chemical elements rates at the end of BBN, which results in an upper limit $\Delta N\mrm{eff} \lesssim 0.5$, as computed with the public code \texttt{AlterBBN}~\cite{Arbey:2011nf,Arbey:2018zfh}. These limits are plotted in Fig.~(6) of~\cite{Arbey:2021ysg}.

\section{Results and discussion}

Here we present some of the results of~\cite{Arbey:2021ysg}. We have computed the contribution to $\Delta N\mrm{eff}$ in three cases: $i)$ a monochromatic spin distribution, $ii)$ the extended spin distribution issued from PBH hierarchical mergers~\cite{Fishbach:2017dwv} and $iii)$ the extended spin distribution issued from PBH formation during an EMDE~\cite{Harada:2017fjm}. The first 2 cases are a refinement of the studies by~\cite{Hooper:2020evu,Masina:2020xhk,Masina:2021zpu} and a detailed comparison can be found in~\cite{Arbey:2021ysg}, which advocates the importance of the precise computation of the reheating time $t\mrm{RH}$ and the functions $g_*$, $g_{*,S}$ at reheating. The third case is original to this study and will be the focus of the present discussion. PBHs formed in an EMDE can be born with a very high spin, depending on two effects: the non-gaussianity of the primordial density fluctuation (denoted as $1^{\rm st}$ order) and the inhomogeneity of the collapsing region (denoted as $2^{\rm nd}$ order). These effects are quantified by the width of the primordial density power spectrum $\sigma\mrm{H}$ and result in spin distributions $\d n^{1,2}/\d a^*$ respectively peaked at high and low spins (see Figs.~(4) and (3) of~\cite{Harada:2017fjm}).

The scenario is the following: an EMDE is caused by the presence of some field in the early universe after inflation; during this EMDE, PBHs form through the collapse of local overdensities; then the field decays and PBHs evolve during a radiation-dominated era; as the PBH density scales as $a^{-3}$ and radiation scales as $a^{-4}$, where $a$ is the expansion factor, if PBHs are initially sufficiently abundant they come to dominate the universe before complete evaporation; PBHs finally evaporate, causing a reheating in the form of SM radiation on top of graviton DR; the latter is probed by BBN and CMB. A realization of this scenario is described in details in the Appendix~A of~\cite{Arbey:2021ysg}.

\begin{figure}
	\centering
	\includegraphics[scale = 1]{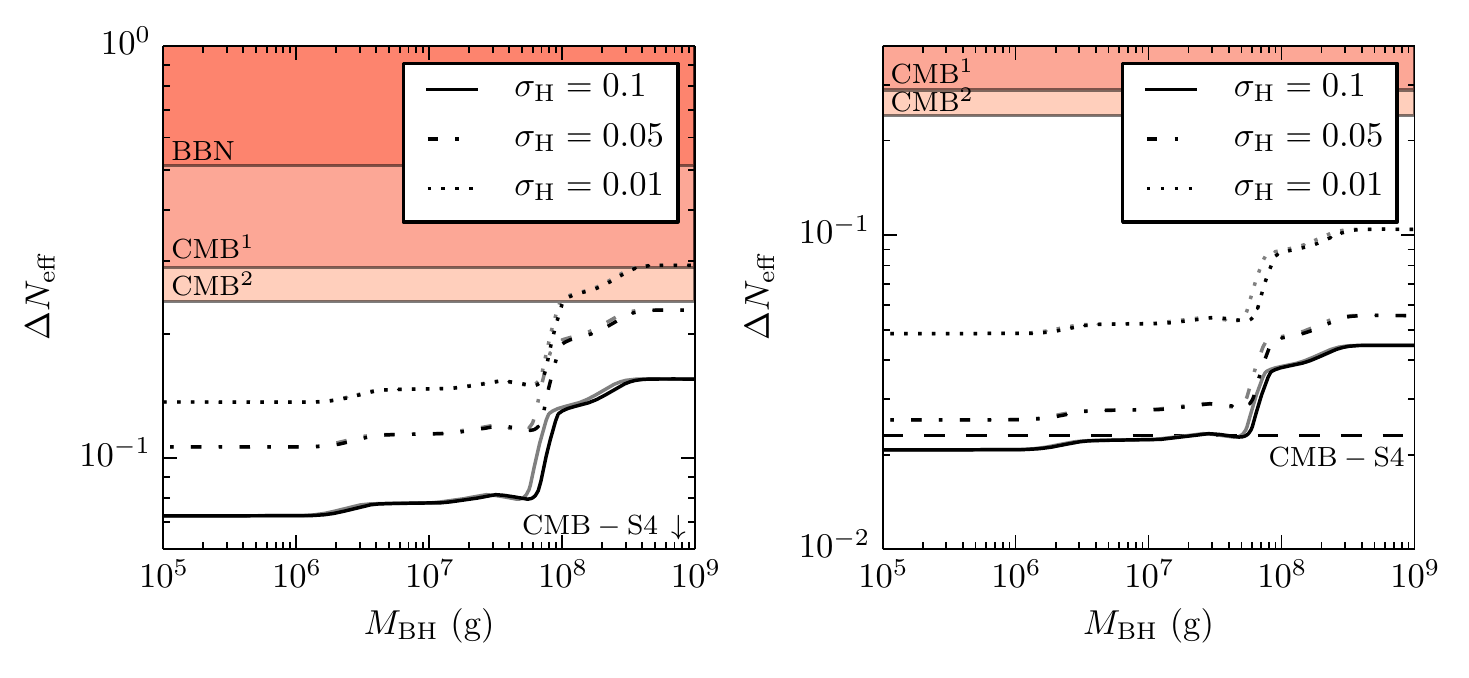}
	\caption{Contributions to $\Delta N\mrm{eff}$ from first (left panel) and second (right panel) order effects, with power spectrum widths $\sigma = \{ 0.1, 0.05, 0.01\}$. All curves are flat below $M < 10^5$, thus the plots are truncated at this mass value. The line colors correspond to the two distinct ways of computing the reheating time: weighted average (black) and last PBH evaporation (grey). The constraints from CMB and BBN are taken from Fig.~(6) of~\cite{Arbey:2021ysg}. \textit{Adapted from~\cite{Arbey:2021ysg}.}\label{fig:1}}
\end{figure}

The results are shown in Fig.~\ref{fig:1}. The two panels correspond to the two kinds of effects. What we conclude in the light of these results is that the high PBH spins of the $1^{\rm st}$ order effect result in plentiful graviton production such that the upper part of the mass range under consideration is already excluded by current CMB limits for $\sigma\mrm{H}\lesssim 10^{-2}$. Both effects will be probed by the future CMB-S4 experiment, which will be able to exclude part or all of the PBH mass range $M = 10^{-5}-10^9\,$g, depending on the value of the power spectrum width $\sigma\mrm{H}$.

\paragraph{Summary} We have seen that the Hawking radiation of hot gravitons from very light PBHs in the early universe, prior to BBN, leads to tight constraints in a previously unconstrained mass range~\cite{Arbey:2021ysg}. If PBHs are born during an EMDE, they have a high initial spin and hence an enhanced rate of graviton emission. Those contribute to $\Delta N\mrm{eff}$, which will be deeply probed by CMB-S4.

\paragraph{Acknowledgment} The work of P.S. and B.S. is supported in part by NSF grant PHY-2014075.

\bibliographystyle{JHEP}
\bibliography{biblio}

\end{document}